\documentclass[twocolumn]{aastex631}

\newcommand{\abund}[1]{$\log N({\rm #1})/N({\rm He})$}

\newcommand{\figref}[1]{Fig.~\ref{#1}}

\newcommand{\hone}{\ion{H}{1}}

\newcommand{\hetwo}{\ion{He}{2}}

\newcommand{\cthree}{\ion{C}{3}}
\newcommand{\cfour}{\ion{C}{4}}

\newcommand{\nthree}{\ion{N}{3}}
\newcommand{\nfour}{\ion{N}{4}}
\newcommand{\nfive}{\ion{N}{5}}

\newcommand{\othree}{\ion{O}{3}}
\newcommand{\ofour}{\ion{O}{4}}
\newcommand{\ofive}{\ion{O}{5}}
\newcommand{\osix}{\ion{O}{6}}

\newcommand{\sifour}{\ion{Si}{4}}

\newcommand{\pfive}{\ion{P}{5}}

\newcommand{\sfour}{\ion{S}{4}}
\newcommand{\sfive}{\ion{S}{5}}

\newcommand{\fefive}{\ion{Fe}{5}}

\newcommand{\ebv}{$E(B-V)$\/}

\newcommand{\kms}{km s$^{-1}$}
\newcommand{\logg}{$\log g$}

\newcommand{\msun}{$M_{\sun}$}

\newcommand{\teff}{$T_{\rm eff}$}
\newcommand{\vhelio}{$v_{\rm Helio}$}
\newcommand{\vinf}{$v_{\infty}$}
\newcommand{\vlsr}{$v_{\rm LSR}$}
\newcommand{\vrot}{$v \sin i$}

\newcommand{\fuse}{{FUSE}}

\newcommand{\hst}{{HST}}

\newcommand{\iue}{{IUE}}

\shorttitle{ZNG~1 in M5}
\shortauthors{Dixon}
\begin{document}

\title{Observations of the Ultraviolet-Bright Star ZNG~1 in the Globular Cluster M5 (NGC~5904)}

\correspondingauthor{William V. Dixon}
\email{dixon@stsci.edu}

\author[0000-0001-9184-4716]{William V. Dixon}
\affiliation{Space Telescope Science Institute, 3700 San Martin Drive, Baltimore, MD 21218, USA}



\begin{abstract}

We have analyzed archival spectra of the hot UV-bright
star ZNG~1 in the globular cluster M5 (NGC~5904) obtained
with the Far Ultraviolet Spectroscopic Explorer (FUSE) and the 
Space Telescope Imaging Spectrograph (STIS).
From these data, we derive 
an effective temperature \teff\ = $43{,}000 \pm 1400$ K,
a surface gravity \logg\ = $4.47 \pm 0.08$, 
a rotational velocity \vrot\ = $157 \pm 12$ \kms, and
a mass $M = 0.92 \pm 0.17$ \msun.  The atmosphere is
helium-rich ($Y = 0.99$) and enhanced in CNO (relative to the cluster).  The spectrum exhibits wind features with a terminal velocity near 1500 \kms\ and strong discrete absorption components (DACs).
The high helium abundance, stellar mass, and rotational velocity suggest that the
star is a merger remnant, and its parameters are consistent with models of a pair of merging He-core white dwarfs.

%% The AAS Journals have a 250 word limit for the abstract.

\end{abstract}

\keywords{stars: abundances --- stars: atmospheres --- stars: individual (\object[Cl* NGC 5904 ZNG 1]{NGC 5904 ZNG~1}) --- ultraviolet: stars}


\section{Introduction} \label{sec:intro}

\citet[][hereafter ZNG]{ZNG:1972} identified seven UV-bright stars---stars that lie above the horizontal branch and to the left of the giant branch in the Hertzsprung--Russell diagram---in the globular cluster M5 (NGC~5904).  The brightest of these, ZNG~1, is a puzzling object with an unusual chemistry and a high rotational velocity.  

The star was first studied by \citet{Bohlin:1983}, who obtained ultraviolet images of the cluster with a rocket-borne telescope.  These images revealed a hot, luminous star some 20\arcsec\ from the cluster center, which the authors identified as ZNG~1.  Follow-up observations with the International Ultraviolet Explorer (IUE) showed strong resonance lines of \nfive\ $\lambda 1240$ and \cfour\ $\lambda 1550$ and emission in the \nfour ] $\lambda 1487$ line.  The authors estimated an effective temperature around 35,000 K.  The \nfive\ $\lambda 1240$ doublet is particularly strong, suggesting that the star is enhanced in nitrogen.  The  \nfour ] emission was interpreted as evidence that a nitrogen-rich planetary nebula (PN) may be present.

Combining several IUE spectra and comparing them to solar-abundance \citet{Kurucz:79} models, \citet{deBoer:1985} derived \teff\ = $28{,}000 \pm 4000$ and $\log g = 4.5$ and argued that the apparent \nfour ] $\lambda 1487$ emission is spurious.

To investigate the possibility of a PN around ZNG~1, \citet{Napiwotzki:97} obtained H$\alpha$ imaging and UV spectroscopy with the Wide Field and Planetary Camera 2 (WFPC2) and the Goddard High Resolution Spectrograph (GHRS), respectively, aboard the Hubble Space Telescope (HST).  The star cannot be observed from the ground, because it lies $0\farcs5$ from a bright G-type companion.  The star is easily resolved from its companion in the WFPC2 H$\alpha$ image, and there is no evidence of a planetary nebula.  The GHRS spectrum exhibits strong wind features in both \nfive\ $\lambda 1240$ and \cfour\ $\lambda 1550$ and photospheric lines of \cthree, \cfour, \nthree, \nfour, \ofour, \ofive, and \sfive.  The spectrum is similar to that of HD~128220~B, for which \teff\ = $40{,}600 \pm 400$ K, \logg = $4.5 \pm 0.1$ \citep{Rauch:93}.

Observations with the Far Ultraviolet Spectroscopic Explorer (FUSE) revealed that the star is most unusual.  By fitting its spectrum with non-LTE (local thermodynamic equilibrium) models, \citet{Dixon:2004} derived \teff\ = $44{,}300 \pm 300$ K, \logg\ = $4.3 \pm 0.1$, and a rotational velocity \vrot\ = $170 \pm 20$ \kms.  The atmosphere is helium-rich and enhanced in carbon, nitrogen, and oxygen.  The spectrum shows evidence of a wind and blue-shifted, highly-ionized absorption, either around the star or along the line of sight.

\citet{Zech:2008} combined the \fuse\/ data with E140M echelle spectra obtained with the Space Telescope Imaging Spectrograph (STIS) on \hst.  They found high-velocity absorption in \cfour, \sifour, \osix, and lower-ionization species with local standard of rest (LSR) velocities of roughly $-140$ and $-110$ \kms.  Using photoionization models and path-length arguments, they conclude that the gas is not circumstellar but represents absorption from highly-ionized high-velocity clouds along the line of sight.  

In this paper, we return to a consideration of the star itself.  Using the full set of \fuse\/ and STIS data (Table \ref{tab:log_obs}) and the latest non-LTE models, we re-derive the stellar parameters, as well as the abundances of C, N, O, Si, S, and Fe.  We model the wind features.  We argue that the star is the product of a stellar merger, perhaps of a pair of He-core white dwarfs.

In Section \ref{sec_observations}, we present our data.  In Section \ref{sec_analysis}, we discuss our atmospheric models and use them to derive stellar parameters and abundances.  In Section \ref{sec_wind}, we model the wind profiles and estimate the mass-loss rate.  In Section \ref{sec_discussion}, we consider possible formation scenarios.  We summarize our conclusions in Section \ref{sec_conclusions}.

%%%%%
% Table 1
\begin{deluxetable*}{lcccclcc}
\tablecaption{Summary of {\it FUSE} and STIS Observations \label{tab:log_obs}}
\tablehead{
\colhead{Instrument} & \colhead{Grating} & \colhead{Wavelength} & \colhead{$R\equiv\lambda/\Delta\lambda$} & \colhead{Exp. Time} & \colhead{Obs.\ Date} & \colhead{Data ID} & \colhead{P.I.} \\
& & \colhead{(\AA)} & & \colhead{(s)}
}
\startdata
\fuse\ & $\cdots$ &\phn905--1187 & 20,000 &   \phn3,734 & 2000 July 15 & A1080303 & Dixon \\
                 &                      &                    &             &   26,761 & 2003 Apr 11-13 & D1570301-3 & Dixon \\
STIS & E140M & 1140--1729 & 45,800 & \phn4,950 & 2003 Jul 08 & O6N404010-20 & Howk \\
                 &                      &                    & & \phn7,850 & 2003 Jul 18-19 & O6N403010-30 & Howk \\
\enddata
\end{deluxetable*}
%%%%%

\section{Observations and Data Reduction}\label{sec_observations}

\subsection{\fuse\/ Data}

\fuse\/ provides medium-resolution spectroscopy from 1187 \AA\ to the Lyman limit \citep{Moos:00, Sahnow:00}.  ZNG~1 was observed through the \fuse\/ $30\arcsec \times 30\arcsec$ aperture.  The data were reduced using v3.2.2 of CalFUSE, the standard data-reduction pipeline software \citep{Dixon:07}, and retrieved from the Mikulski Archive for Space Telescopes (MAST).  The \fuse\/ data used in this work are available at \dataset[10.17909/158e-8q57]{https://doi.org/10.17909/158e-8q57}. For each \fuse\/ channel, the extracted spectra from all exposures were shifted to a common wavelength scale, weighted by exposure time, and combined into a single file. 

The signal-to-noise (S/N) ratio of the resulting spectrum ranges from 10 to 20 per 0.05 \AA\ resolution element in the SiC bands (900--1000 \AA) and from 30 to 40 in the LiF bands (1000-1187 \AA).  In most of our model fits, we employ only the spectrum from the channel with the highest S/N ratio.  An exception is the 900--1000 \AA\ region, for which we have combined data from the SiC 1B and SiC 2A detectors, weighted by exposure time, into a single spectrum.  The two channels have different line-spread functions, so combining their spectra reduces the spectral resolution; however, this effect is insignificant given the star's high rotational velocity.

The \fuse\/ wavelength calibration is reasonably accurate, but small offsets of the target within the $30\arcsec \times 30\arcsec$ aperture can introduce a zero-point offset in the wavelength scale.   To place the data on an absolute wavelength scale, we shift the spectrum from each channel so that the velocities of their interstellar lines match those of our model, described in Section \ref{sec_ism}.

\subsection{STIS Data}
\label{STIS}

The design and construction of STIS are described by \citet{Woodgate:STIS:1998}, and information about its on-orbit performance is provided by \citet{Kimble:STIS:1998}.  ZNG~1 was observed using the STIS far-ultraviolet MAMA detector with the E140M grating and the $0\farcs2 \times 0\farcs2$ aperture.  The data were retrieved from StarCAT \citep{Ayres:StarCAT:DOI, Ayres:StarCAT:2010}, a collection of high-resolution ultraviolet STIS spectra hosted at MAST.  The StarCAT project combined data from all observations, exposures, and spectral orders of every star observed with STIS during its first seven years on orbit into a single, fully-calibrated spectrum.  The STIS calibration pipeline yields spectra with a heliocentric wavelength scale; StarCAT spectra have an absolute wavelength uncertainty of approximately 1.4 \kms.

In the course of our analysis, we discovered that the flux error bars in the StarCAT spectrum are too large by about a factor of two.  For example, in a flat region of the spectrum near 1172 \AA, the standard deviation of the flux array is 0.08 (in arbitrary units), while the mean value of the error array is 0.15.  Repeating this calculation across the STIS spectrum, we derive an average correction factor of 0.55 for the error array.  Applying this correction, we find that the StarCAT spectrum of ZNG~1 has a S/N ratio ranging from about 20 to 50 per two-pixel resolution element.

\section{Analysis}\label{sec_analysis}

\subsection{Interstellar Medium}\label{sec_ism}

The FUV spectrum of ZNG~1 is punctuated by a variety of interstellar absorption features.  In the \fuse\/ bandpass, molecular hydrogen is the dominant species, while high-ionization species are prominant at longer wavelengths.  \citet{Zech:2008} concluded that the high-velocity absorption from both high- and intermediate-ionization species is due to high-velocity clouds along the line of sight.  We use the \citeauthor{Zech:2008} results to model the interstellar absorption,  with high-velocity components at \vlsr\ $\sim \; -140$ and $-110$ \kms\ and low-velocity components at \vlsr\ $\sim \; -59$ and $-6$ \kms.  \citeauthor{Zech:2008} report that a two-component model, with \vlsr\ = $-135$ and $-18$, fits the \hone\ absorption as well as the four-component model, so we adopt the simpler model for this species.  We convert to a heliocentric reference frame using the relation \vlsr\ $-$ \vhelio\ = $+13.25$ \kms.  For low-velocity species not fit by \citeauthor{Zech:2008}, we fit the absorption by eye, which is sufficient for our needs.

Synthetic interstellar absorption spectra are computed using software written at the University of California, Berkeley, by M. Hurwitz and V. Saba. Given the column density, Doppler broadening parameter, and velocity of each component, the program computes a Voigt profile for each absorption feature and produces a high-resolution spectrum of optical depth versus wavelength.  Wavelengths, oscillator strengths, and other atomic data are taken from \citet{Morton:03}.

\subsection{Model Atmospheres}\label{sec_models}

We compute non-LTE stellar-atmosphere models using version 208 of the program TLUSTY \citep{Hubeny:Lanz:95}.  We employ atomic models similar to those used by \citet{Lanz:Hubeny:2003} to compute their grid of O-type stars.  Within TLUSTY, one normally specifies all chemical abundances relative to hydrogen; however, for hydrogen-deficient objects, \citet{Hubeny:Lanz:2017c} recommend the use of helium as the reference atom, and we have taken this advice.  Since CNO are the most abundant metals in the star's photosphere, we begin by generating models with H+He+C+N+O.  To determine the carbon abundance, we fix N and O at the levels determined by \citet{Dixon:2004} and generate a grid of models that vary in C/He.  We repeat the process for N and O.  For all other species, we generate a grid of models with H+He+C+N+O+X, where CNO are fixed at their best-fit values.

Given a model atmosphere, we compute a synthetic spectrum using version 54 of the program SYNSPEC \citep{Hubeny:88}.  For the \fuse\/ data, the synthetic spectra are convolved with a Gaussian of FWHM = 0.06 \AA\ to match the \fuse\/ line-spread function.  For the STIS spectrum, we use a Gaussian with FWHM = 0.025 \AA.  We multiply the synthetic spectrum by a \citet{Fitzpatrick:1999} extinction curve assuming \ebv\ = 0.03 \citep{Harris:96, Harris:2010}, extrapolated to the Lyman limit, and the ISM absorption model described above.  Finally, we scale the model to reproduce the continuum in a nearby (apparently) line-free region in the observed spectrum.

Given a grid of synthetic spectra, our fitting routine linearly interpolates among them---in one, two, or three dimensions, as appropriate---determining the best fit to the data via chi-squared minimization.  The uncertainties quoted for parameters derived from individual line fits are $1 \sigma$ errors computed from the covariance matrix returned by the fitting routine; we refer to these as statistical errors.

Continuum placement is the dominant uncertainty in our fits.  Because of its high rotational velocity, only the strongest features in the spectrum of ZNG~1 can be identified unambiguously.  Weak features from additional species may depress the apparent continuum.  Allowing our fitting routines to scale the model to the mean level of the ``pseudo-continuum'' thus underestimates the true continuum level.  To estimate the uncertainty inherent in our continuum estimate, we perform each fit twice, once with the model continuum fixed as described and again with the model scaled by a factor of 0.97.  The difference in the two abundances is an estimate of the systematic error in our abundance estimates. We add this term and the statistical error in quadrature to compute our final error for a single absorption feature.  In most cases, the continuum uncertainty is the dominant contributor to the final error.

\citet{Dixon:2004} derived a surface gravity \logg\ = 4.3 and a helium abundance of 85\% by number.  Preliminary analyses of these new data suggest that the actual values are somewhat higher, so we assume \logg\ = 4.5 and $N$(H)/$N$(He) = 0.03 in our initial models.

\subsection{Radial and Rotational Velocities}\label{sec_rv}

To determine the star's radial and rotational velocities, we use the \cthree* multiplet at 1175 \AA, a strong feature uncontaminated by interstellar absorption.  The line appears in both the \fuse\/ and STIS spectra.  We begin with the STIS spectrum, which has higher S/N and spectral resolution.  We generate models with carbon abundance 1, 3, and 6 times some fiducial value.  For each model, we generate synthetic spectra with rotational velocities between 120 and 200 \kms\ in steps of 10 \kms.  Assuming a radial velocity \vhelio\ = 30 \kms, we fit the spectrum, allowing the carbon abundance and rotational velocity to vary freely, then compute $\chi^2$ for the best-fit model.  We repeat the process for integral values of \vhelio\ between 30 and 60 \kms.  We find that the STIS spectrum is best reproduced by a model with \vhelio\ = $42 \pm 2$ \kms\ and \vrot\ = $150 \pm 15$ \kms.  The error bar on the heliocentric velocity represents the range of velocities whose $\chi^2$ values lie within $\Delta \chi^2 = 2.3$ of the minimum (the range appropriate for two parameters of interest; \citealt{Press:89}); it is thus purely statistical.  The error bar on the rotational velocity is computed as described above.  We repeat the process for the \fuse\/ spectrum, deriving \vhelio\ = $45 \pm 3$ \kms\ and \vrot = $168 \pm 19$.  

We adopt the weighted mean of the \fuse\/ and STIS values, \vhelio\ = $43 \pm 2$ \kms\ and \vrot\ = $157 \pm 12$ \kms\  (where we weight by the measurement uncertainties), as the radial and rotational velocities  of ZNG~1. The cluster has a mean heliocentric velocity of $53.2$ \kms, with a central velocity dispersion of 5.5 \kms\ \citep{Harris:96, Harris:2010}, so the velocity of ZNG~1 is within $\sim 2 \sigma$ of the cluster mean.

\begin{figure}
\epsscale{1.2}
\plotone{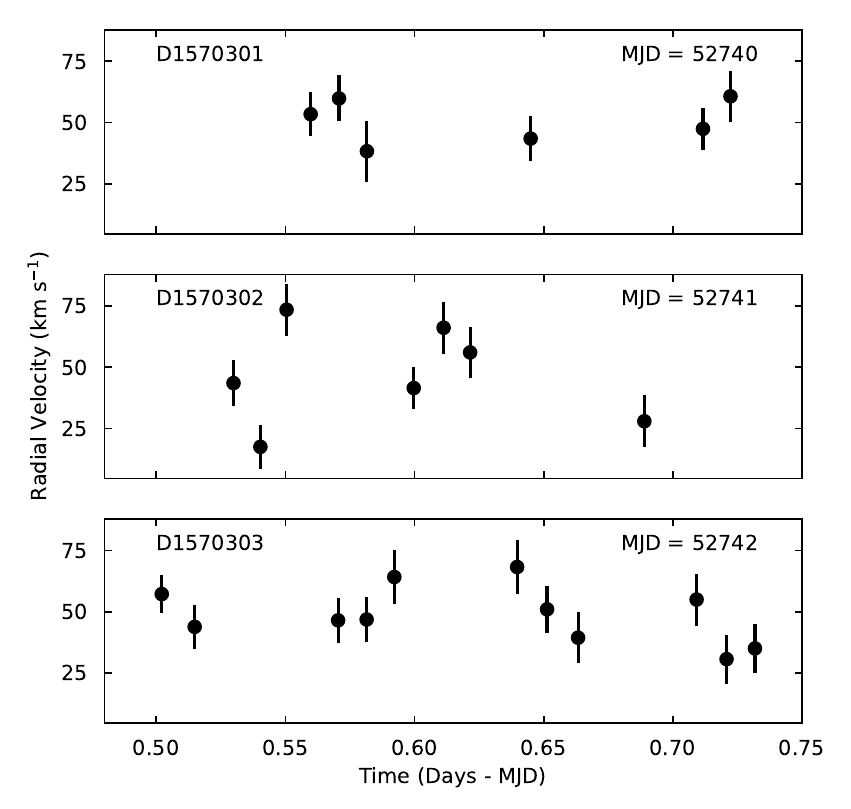}
\caption{Radial velocities derived from the \fuse\/ observations of ZNG~1 in M5.}
\label{fig_velocities}
\end{figure}

\citet{Lanz:2004} derived a rotational velocity \vrot\ = 100 \kms\ from the \fuse\/ spectrum of the He-rich sdB star PG~1544+488.  From a high-resolution optical spectrum, \citet{Ahmad:2004} determined that the star is actually a spectroscopic binary.  Because the \fuse\/ observation was obtained in time-tag mode, the authors were able to divide the data into short sub-exposures, extract a spectrum from each, and show that the best-fit radial velocity varies over the course of several hours.  To see if a similar effect is at work in ZNG~1, we divide the three \fuse\/ D15703 observations into sub-exposures roughly 1000 s long, place the resulting spectra on a heliocentric wavelength scale by shifting them to match our ISM model between 1120 and 1160 \AA, and fit a Gaussian to the \cthree* multiplet at 1175 \AA. The resulting radial velocities are plotted in \figref{fig_velocities}.  Error bars are computed from the covariance matrix and are thus purely statistical.  The radial velocities derived from observations D1570301 and D1570303 show some scatter, but are generally consistent with a constant value.  The velocities derived from observation D1570302 show more scatter, but no significant trend.  We conclude that the \fuse\/ data provide no evidence that the star is a spectroscopic binary and proceed under the assumption that its broad absorption features reflect its high rotational velocity.  

The STIS spectra have both higher resolution and a higher S/N ratio; unfortunately, they were obtained in ACCUM mode, so are not amenable to this analysis.

\begin{figure}
\epsscale{1.2}
\plotone{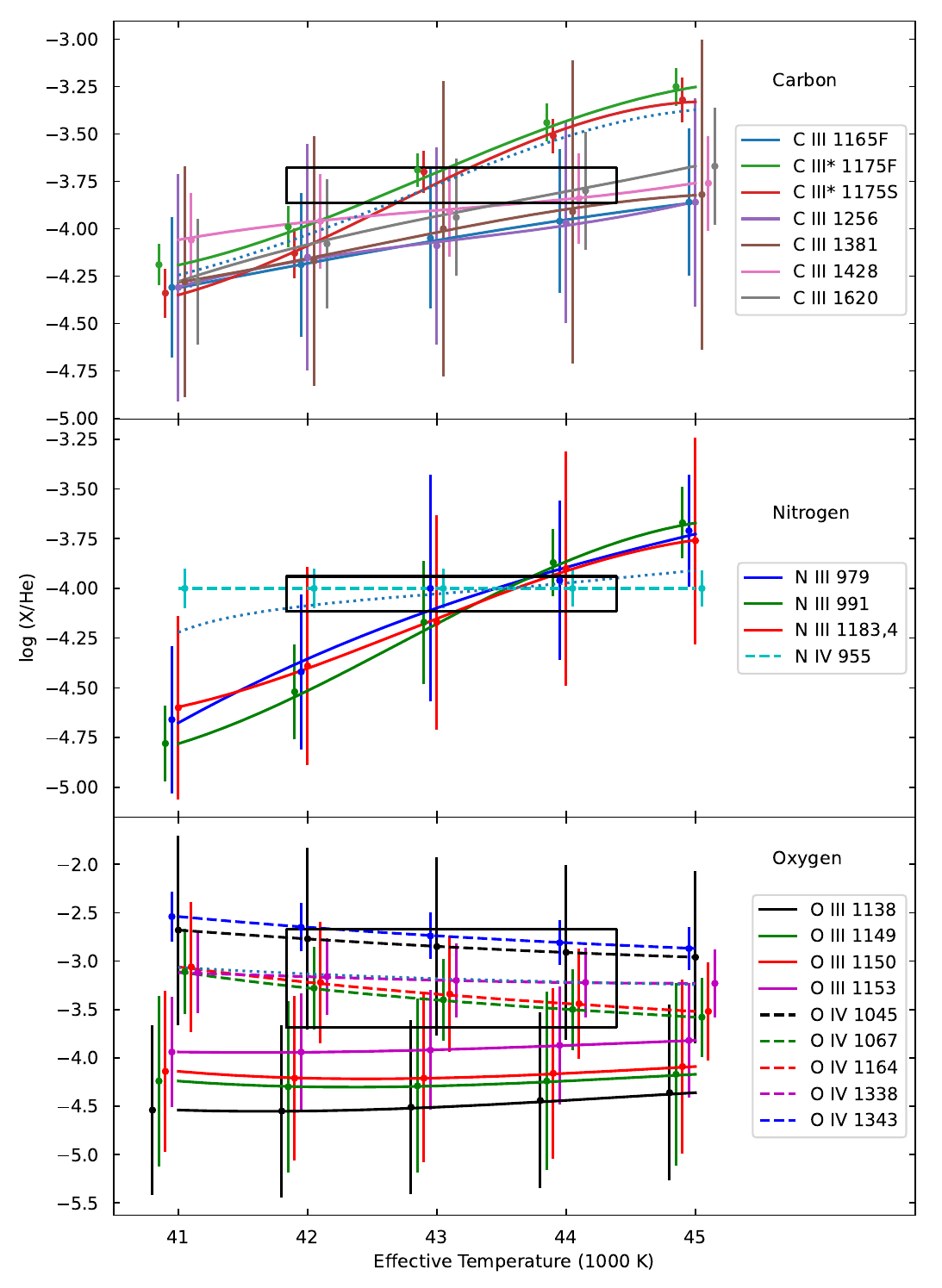}
\caption{Deriving the effective temperature and abundance of C, N, and O.  Points with error bars represent the abundance derived from model fits to each absorption feature.  Solid and dashed lines are low-order polynomial fits to the measured points.  Dotted lines represent the weighted mean of the measured abundances, computed at 10 K intervals.  The black box in each panel denotes the allowed ranges of temperature and abundance, as described in the text.}
\label{fig_teff}
\end{figure}

\subsection{Effective Temperature and CNO Abundances}\label{sec_teff}

Following \citet{Dixon:2019}, we use the absorption features of multiple ionization states of CNO to derive the star's effective temperature.  Consider the middle panel of \figref{fig_teff}.  Using a series of models with \teff\ = 41,000 K and \abund{N} between $-5.0$ and $-3.1$, we fit the \nthree\ $\lambda 979$ feature to determine the nitrogen abundance.  We repeat using models with temperatures increasing in steps of 1000 K to 45,000 K.  The resulting nitrogen abundances are plotted as dark blue points and connected by a low-order polynomial (evaluated at 10 K intervals).  Vertical bars represent the uncertainties returned by the fitting routine.  As the temperature rises, the fraction of nitrogen in the form of \nthree\ falls, requiring a higher nitrogen abundance to reproduce the observed feature.  We repeat this procedure for the \nthree\ $\lambda 991$ line, the \nthree\ $\lambda \lambda 1183, 1184$ doublet, and the strong \nfour\ feature at 955 \AA.  Interestingly, the strength of the \nfour\ line is constant in this temperature regime.

We repeat the process for C and O (top and bottom panels of \figref{fig_teff}).  Note that we have excluded from consideration features that yield wildly discrepant abundances, as well as those that yield abundances with error bars so large that they provide no useful constraints.  This filter eliminates all of the \cfour\ lines\footnote{For example, for models with \teff\ = 43,000 K, the \cfour\ $\lambda \lambda 1351, 1353$ doublet yields \abund{C} = $-2.49 \pm 0.38$; the \cfour\ $\lambda \lambda 1548, 1550$ resonance doublet, which is blended with a strong wind feature, yields an upper limit of \abund{C} $< -3.30$; and a broad, weak \cfour\ absorption feature centered at 1654 \AA\ yields \abund{C} = $-3.40 \pm 1.38$.}, leaving only \cthree\ and \cthree*.   The \cthree* $\lambda 1175$ multiplet appears in both the \fuse\/ and STIS spectra, so we fit it twice.  We see that the curves for these two species cross at temperatures near 41,000 K, while the nitrogen curves cross at temperatures near 44,000 K.  The \othree\ and \ofour\ curves remain stubbornly separate.  To combine these results in a quantifiable way, we compute the error-weighted mean abundance and the error-weighted standard deviation as a function of temperature for each element.  The mean abundance is plotted as a dotted line in each panel.  At each temperature step, we compute
$$\chi^2 = \sum \Bigl \lbrace [y_i - y(x_i)]^2 / \sigma_i^2 \Bigr \rbrace,$$
where $y_i$ is the abundance derived from a single feature (or group of features; solid or dashed line), $y(x_i)$ is the mean abundance (dotted line), $\sigma_i$ is the uncertainty in the derived abundance (vertical bars), and the summation is taken over the 20 abundance curves plotted in \figref{fig_teff}.  $\chi^2$ has a minimum at \teff\ = 43,000 K.  Our upper limit to the temperature is \teff\ = 44,400 K, set by the point at which $\chi^2$ rises by 4.72 relative to its minimum.  This choice of $\Delta \chi^2$ is strictly correct for a model with four interesting parameters (one temperature and three abundances) if all errors are normally distributed \citep{Press:89}.  For each element, our best-fit abundance is the error-weighted mean value (computed above) at the best-fit temperature.  The abundance uncertainty is the larger of the error-weighted standard deviation or the uncertainty in the weighted mean.  The black box in each panel denotes the allowed range of temperature and abundance for that element.  Derived abundances are presented in Table \ref{tab:abundance}.

\subsection{Surface Gravity and Hydrogen Abundance}\label{sec_hydrogen}

\begin{figure}
\epsscale{1.2}
\plotone{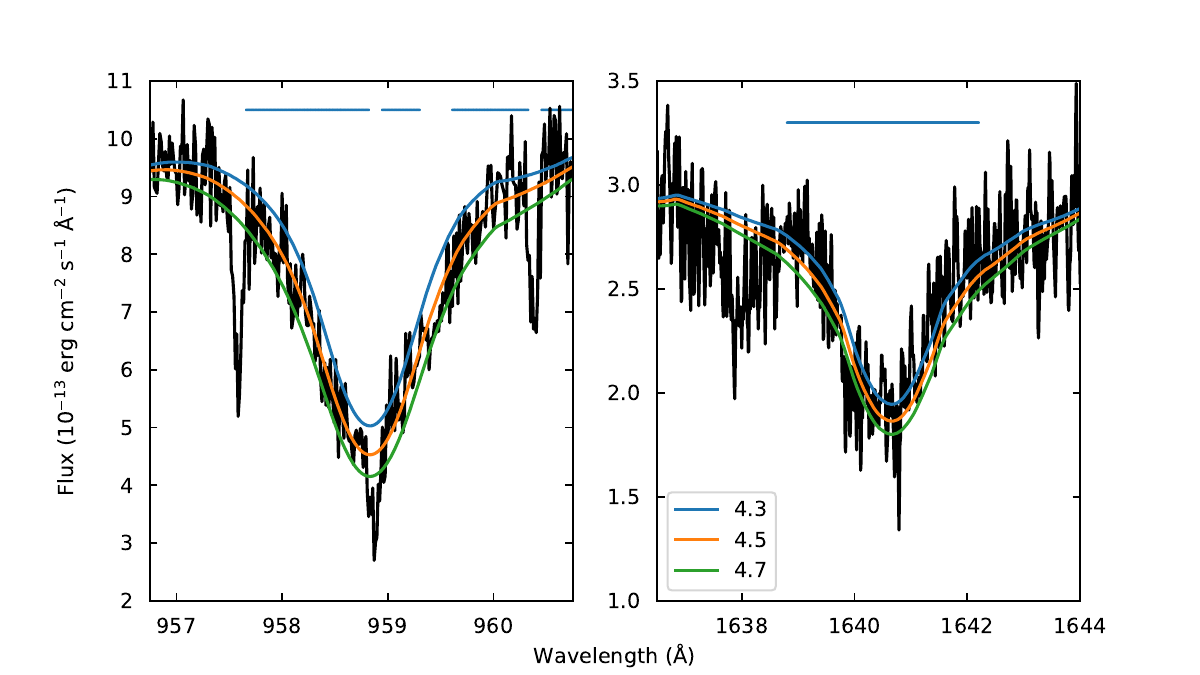}
\caption{\hetwo\ $\lambda 958$ (left) and $\lambda 1640$ (right) features in the spectrum of ZNG~1 in M5.  Synthetic spectra with surface gravity \logg\ = 4.3, 4.5, and 4.7 are overplotted.  Blue bars indicate the spectral regions included in our fit.}
\label{fig_1640}
\end{figure}

As shown in \figref{fig_1640}, the strong \hetwo\ features at 958 and 1640 \AA\ provide a sensitive probe of the surface gravity.  Assuming an effective temperature of \teff\ = 43,000 K, we fit each line separately.  The 958 \AA\ line in the \fuse\/ spectrum yields a surface gravity \logg\ = $4.51 \pm 0.09$, while the 1640 \AA\ line in the STIS spectrum yields \logg\ = $4.37 \pm 0.16$.  The weighted mean of these two values is \logg\ = $4.48 \pm 0.08$.

\begin{figure}
\epsscale{1.2}
\plotone{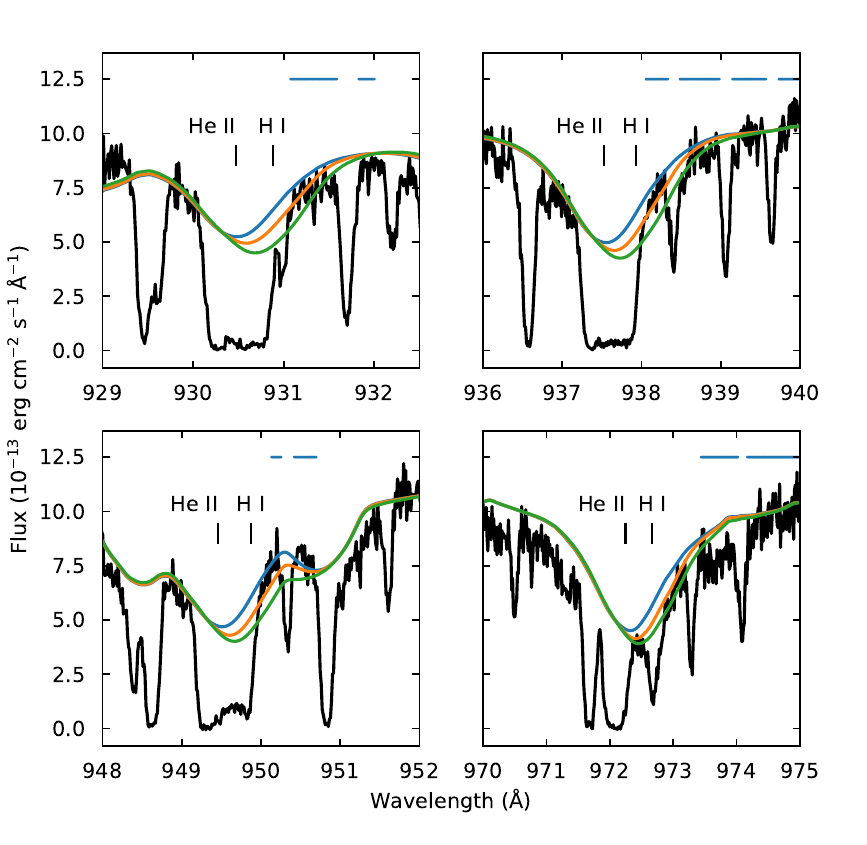}
\caption{Hydrogen and helium lines in the \fuse\/ spectrum of ZNG~1 in M5.  The narrow lines are interstellar.  Synthetic spectra with $N$(H)/$N$(He) ratios of 0.05 (green), 0.01 (orange) and 0.001 (blue) are overplotted.  An ISM model, which is incorporated into our models, is omitted from this figure for clarity.  Blue bars indicate the spectral regions included in our fits.}
\label{fig_hydrogen}
\end{figure}

%%%%%
% Table 3
\begin{deluxetable}{lcc}
\caption{Photospheric Abundances \label{tab:abundance}}
\tablehead{
\colhead{Species} & \colhead{$\log N({\rm X})/N({\rm He})$} & \colhead{$\log \beta_{\mathrm X}$\tablenotemark{a}}
}
\startdata
H & $-1.70 \pm 0.30 $ & $-2.30 \pm 0.30$ \\
He & \nodata & $-0.00371$ \\
C  & $-3.77 \pm 0.09$ & $-3.30 \pm 0.09$ \\
N  & $-4.03 \pm 0.09$ & $-3.49 \pm 0.09$ \\
O  & $-3.18 \pm 0.51$ & $-2.58 \pm 0.51$ \\
Si & $-5.81 \pm 0.24$ & $-4.97 \pm 0.24$ \\
S  & $-7.65 \pm 0.86$ & $-6.75 \pm 0.86$ \\
Fe & $-5.84 \pm 0.73$ & $-4.70 \pm 0.73$ \\
\enddata
\tablenotetext{a}{Logarithm of mass fraction.}
\end{deluxetable}
%%%%%

\figref{fig_hydrogen} shows a series of \hone\ Lyman lines, each blended with a \hetwo\ Balmer line.  The flat-bottomed features are due to interstellar \hone; they are shifted to the blue by 31 \kms, while the stellar spectrum is shifted to the red by 43 \kms.  Most of the other ISM features are due to molecular hydrogen.  The colored curves are synthetic spectra with $N$(H)/$N$(He) ratios between 0.05 (green) and 0.001 (blue).  An ISM model, included in our fits, is omitted from this figure for clarity.  Though only the red wing of each line is available, it is clear from the figure that the stellar hydrogen abundance is quite low.  Our fits yield a ratio $N$(H)/$N$(He) = $0.020 \pm 0.015$.  

Because a change in the hydrogen fraction (which is effectively a change in the helium abundance) could affect the strength of the helium lines and thus the derived surface gravity, we repeat our fits of the \hetwo\ features at 958 and 1640 \AA\ using models with a hydrogen abundance $N$(H)/$N$(He) = 0.02.  The derived surface gravity falls by only 0.01 dex, to \logg\ = $4.47 \pm 0.08$, so further iterations are unnecessary.  Because the surface gravity and hydrogen abundance have changed so little from our initial values, we need not adjust the effective temperature and CNO abundances that we derived in Section \ref{sec_teff}.  Our final stellar parameters are presented in Table \ref{tab:stellar_parms}.

%%%%%
% Table 4
\begin{deluxetable}{lcccc}[t]
\tablecaption{Stellar Parameters \label{tab:stellar_parms}}
\tablehead{
\colhead{Parameter} & \colhead{Value}
}
\startdata
\teff\ (K) & $43{,}000  \pm 1400$ \\
\logg\ [cm s$^{-2}$] & $4.47 \pm 0.08$ \\
$R_*/R_{\sun}$ & $0.92 \pm 0.02$ \\
$M_*/M_{\sun}$ & $0.92 \pm 0.17$ \\
$\log (L_*/L_{\sun})$ & $3.42 \pm 0.06$ \\
\vhelio\  (\kms) & $43 \pm 2$ \\
\vrot\ (\kms) & $157 \pm 12$ \\
\enddata
\end{deluxetable}
%%%%%

\subsection{Metal Abundances}\label{sec_abundance}

Four other metals, Si, P, S, and Fe, exhibit absorption lines in the spectrum of ZNG~1, but only of a single ionization state.  We compute their abundances as for the lighter elements.  A list of absorption features and the abundance derived from each is presented in Table \ref{tab:lines_fuv}.  The full set of abundances for ZNG~1 is presented in Table \ref{tab:abundance}.  Notes on individual species follow.

{\em Silicon:}\/ While there are several strong silicon features in our spectrum, most are blended with other photospheric lines.  An exception is the \sifour $\lambda \lambda 1393, 1402$ resonance doublet.  The blue wing of each is contaminated by ISM features, so we fit only the red wing.

{\em Phosphorus:}\/ Our \fuse\/ spectrum includes the \pfive\ resonance transitions at 1117 and 1128 \AA.  The 1117 \AA\ line exhibits what could be a P Cygni profile, but the 1128 \AA\ line does not.  Meanwhile, the 1128.0 \AA\ line is blended with strong \sifour\ $\lambda 1128.3$ and \sfive\ $\lambda 1128.7$ features, which are not well reproduced by our models.  We are thus unable to constrain the star's phosphorus abundance.

{\em Sulfur:}\/  Most of the strong sulfur lines in our spectra are blended with other photospheric features.  Only \sfour\ $\lambda 1072$ appears to be uncontaminated; it is the source of our abundance determination.

{\em Iron:}\/  There are many \fefive\ lines in these spectra, but none are very strong.  Instead of fitting individual features, we fit the region between 1373 and 1381 \AA, where our models predict that iron is the dominant source of absorption.  

The large uncertainties in the quoted abundances of S and Fe reflect the change in the derived abundance when the stellar continuum is scaled by a factor of 0.97.

\begin{deluxetable*}{lcDrcc}
\tablecaption{Selected Metal Lines \label{tab:lines_fuv}}
\tablehead{
\colhead{Ion} & \colhead{$\lambda_{\rm lab}$} & \multicolumn2c{$\log gf$} & \colhead{$E_l$} & \colhead{Abundance} & \colhead{Grating} \\
\colhead{} & \colhead{(\AA)} & \multicolumn2c{} & \colhead{(cm$^{-1}$)}
}
\decimals
\startdata
\sifour  &  1393.755  &  0.03  &  0.000  & $ -5.82 \pm 0.30 $ & E140M \\
\sifour  &  1402.770  &  -0.28  &  0.000  & $ -5.80 \pm 0.40 $ & \\
\sfour  &  1072.996  &  -0.83  &  951.100  & $ -7.65 \pm 0.86 $ & \fuse \\
\fefive &    1373.589   &    0.31  &   187719.001   & $ -5.84 \pm 0.73 $ & E140M \\
        &    1373.679   &    0.19  &   187157.505   & \nodata & \\
        &    1373.964   &   -0.18  &   213649.212   & \nodata & \\
        &    1374.119   &    0.40  &   233848.909   & \nodata & \\
        &    1374.253   &   -1.27  &   216779.102   & \nodata & \\
        &    1374.788   &    0.30  &   258628.499   & \nodata & \\
        &    1375.416   &   -1.00  &   258628.499   & \nodata & \\
        &    1375.790   &   -0.73  &   216860.399   & \nodata & \\
        &    1376.337   &    0.44  &   188395.286   & \nodata & \\
        &    1376.451   &    0.05  &   186725.501   & \nodata & \\
        &    1377.718   &   -0.66  &   214525.796   & \nodata & \\
        &    1378.088   &    0.05  &   258769.503   & \nodata & \\
        &    1378.561   &    0.43  &   205536.398   & \nodata & \\
        &    1379.040   &   -0.15  &   209523.911   & \nodata & \\
        &    1379.202   &   -1.12  &   213649.212   & \nodata & \\
        &    1380.112   &   -0.14  &   186433.588   & \nodata & \\
\enddata
\tablecomments{Abundance relative to helium: \abund{X}.}
\end{deluxetable*}

\begin{figure}
\epsscale{1.2}
\plotone{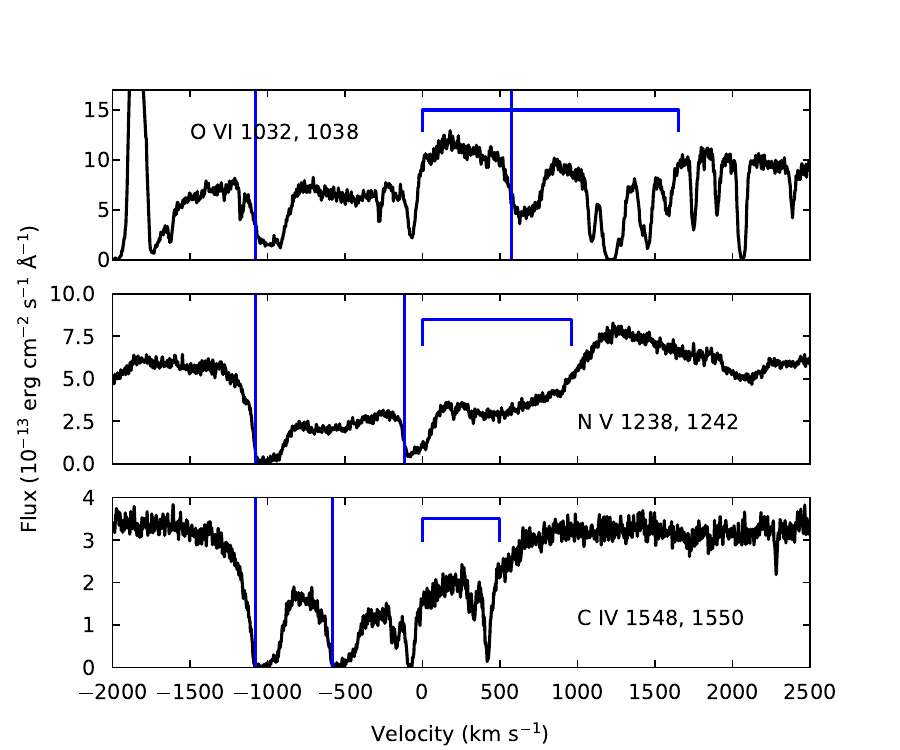}
\caption{Resonance doublets of \osix, \nfive, and \cfour\ in the \fuse\/ and STIS spectra of ZNG~1 in M5, which show strong P Cygni profiles and discrete absorption components (DACs) blueshifted by about 1000 \kms.  Velocities are relative to the short-wavelength component of each doublet.  Horizontal bars mark the positions of the photospheric features.  Vertical bars mark the blue edges of the DACs.  The narrow spectral features are interstellar.}
\label{fig_winds}
\end{figure}

\section{Stellar Wind}\label{sec_wind}

Several absorption features in the FUV spectrum of ZNG~1 in M5 show evidence of a stellar wind.
The P Cygni profile of the \nfive\ $\lambda \lambda 1238, 1242$ doublet was first seen in the star's \iue\ spectrum by \citet{deBoer:1985}, and the star's GHRS spectrum reveals \cfour\ $\lambda \lambda 1448, 1450$ features that are blueshifted by $\sim 900$ \kms\  \citep{Napiwotzki:97}.  In our data, spectacular wind features are present in the resonance doublet of \osix\ $\lambda \lambda 1032, 1038$ as well as those of \nfive\ and \cfour.  These profiles also exhibit discrete absorption components (DACs) blueshifted by about 1000 \kms.  \ofive\ $\lambda 1371$, which is not a resonance line, also exhibits a P~Cygni profile.

\figref{fig_winds} compares the \osix, \nfive, and \cfour\ profiles.  The three profiles are aligned in velocity such that the short-wavelenth photospheric component is at rest.  The DACs have sharp blue edges at $-1080$ \kms, but the absorption extends an additional 400 \kms\ to the blue; this extension is particularly obvious in the \cfour\ profile.

\begin{figure}
\epsscale{1.2}
\plotone{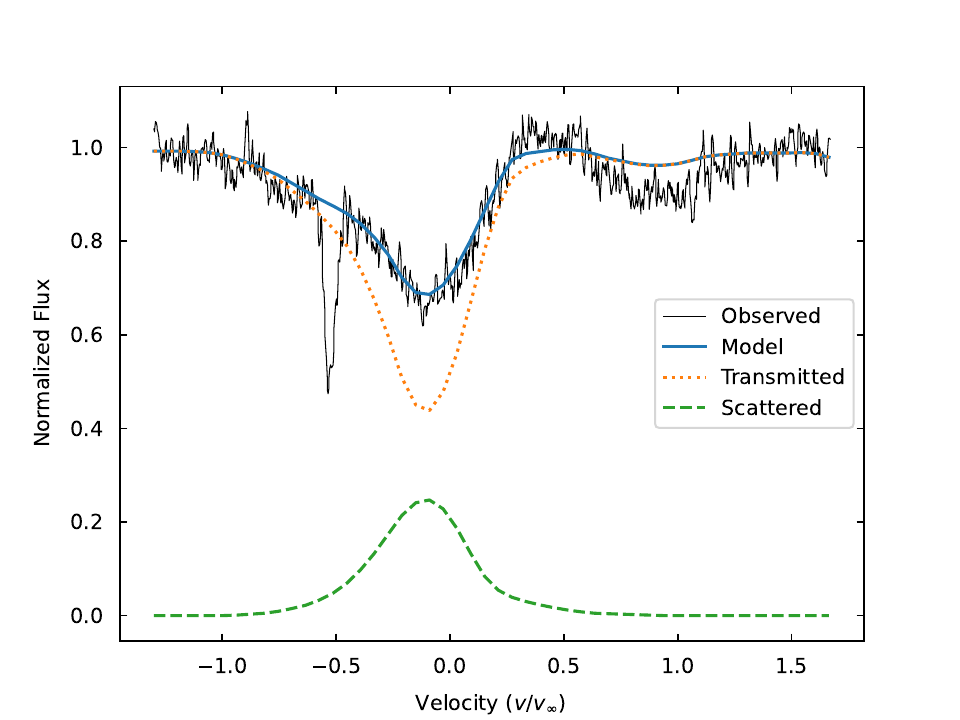}
\caption{Model fits to the \ofive\ $\lambda 1371$ feature, which shows a P~Cygni profile.  The data are overplotted by the wind model described in the text (thick solid line).  Also shown are the transmitted (dotted line) and scattered (dashed line) components of the model profile.  The spectra are plotted in the stellar rest frame.  The transmitted spectrum includes our photospheric absorption model.}
\label{fig_ofive}
\end{figure}

To constrain the properties of the wind, we fit a simple model to the \ofive\ $\lambda 1371$ feature, which shows a P Cygni profile, but no DACs (\figref{fig_ofive}).  We use the line-fitting routines of D.\ Massa, which combine the Sobolev with Exact Integration (SEI) algorithm described by \citet{Lamers:87} with the wind optical depth law introduced by \citet{Massa:95}.  This approach employs the standard beta-law parametrization of the velocity, which has the form 

\begin{equation}
w(x) = w_0 + (1 - w_0) (1 - 1/x)^{\beta},
\end{equation}
where $w(x) = {v(x)}/{v_{\infty}}$, $x = r/R_*$, $R_*$ is the stellar radius, and $w_0$ is the initial velocity at the base of the wind.
(In this formalism, all wind velocities are expressed as a fraction of the terminal velocity \vinf.)  We follow \citet{Massa:95} in adopting $w_0 = 0.01$.  The parameter $\beta$, which defines the steepness of the spatial velocity gradient in the wind, is constrained by the shape of the emission component of the P Cygni line.  Other inputs to the code are the turbulent velocity $w_{\rm D}$, which simulates macroscopic velocity fields in the wind by smoothing the wind profile; the synthetic photospheric spectrum; and  $\tau_{\rm r}(w)$, the radial component of the optical depth of the absorbing species as a function of the wind velocity.

The line profile is fit by adjusting $v_{\infty}$, $\beta$, $w_D$ and the elements of $\tau_{\rm r}(w)$.  The \ion{O}{5} profile is well fit by the parameters $v_{\infty} = 600$ \kms,  $\beta = 0.3$, and $w_{\rm D} = 0.15$.  The optical depth falls smoothly from $\tau = 0.4$ at the base of the wind to zero at $v_{\infty}$.  Our best-fit model is presented in \figref{fig_ofive}.  A complete explanation of this technique, its applicability, and its limitations may be found in \citet{Massa:03}.

From the stellar abundances and our wind model, we can estimate $\dot{M}q(w)$, the product of the stellar mass-loss rate and the fraction of oxygen in the form of O$^{+4}$ as a function of velocity in the wind.  Over the velocity range $0.1 < w < 0.9$, where the model is best constrained, $\dot{M}q(w)$ falls from roughly $2 \times 10^{-12}$ to $3 \times 10^{-14}$ \msun\ yr$^{-1}$.  Our  model predicts that $q = 10^{-3}$ at the outer limit of the stellar atmosphere.  If this value is appropriate for the base of the wind, then $\dot{M} \sim  2 \times 10^{-9}$ \msun\ yr$^{-1}$.  

The blue-shifted absorption component of the \ofive\ $\lambda 1371$ profile extends to about $-600$ \kms, but the corresponding features of the resonance lines extend to $-1500$ \kms.  The \osix\ absorption features are weak in the photospheric spectrum, but the \osix\ wind feature is quite prominent.  Indeed, our models predict that the fraction of oxygen in the form of O$^{+5}$ is less than $10^{-13}$ at the edge of the stellar atmosphere.  We conclude that most of O$^{+4}$ that leaves the star is ionized to O$^{+5}$ before the wind velocity exceeds 600 \kms.

Discrete absorption components (DACs) are ubiquitous in the spectra of luminous OB stars. They are localized (in velocity) absorption features that generally appear at intermediate velocities and accelerate toward higher velocities (that is, to the blue), with a corresponding decrease in velocity width. Attributed to slowly-moving density enhancements in the wind, DACs in O-type stars can represent as much as 50\% of the wind material, although 20\% is more common. A strong correlation between the acceleration and recurrence of DACs and the projected rotation velocity, $v \sin i$, has been observed, though the nature and origin of this connection remain unclear. (See \citealt{Massa:95}, \citealt{Prinja:02}, and references therein.) Given the high rotational velocity of ZNG~1, DACs may be a common feature of its spectrum. Indeed, the two \fuse\/ observations of the star were taken more than three years apart, yet the DACs in their spectra are identical.

\section{Discussion}\label{sec_discussion}

\subsection{Cluster Membership}\label{sec_membership}

Before comparing ZNG~1 with other members of M5, we should confirm that it is a cluster member.  The \hst\/ UV Globular Cluster Survey \citep[HUGS;][]{Piotto:2015, Nardiello:2018} provides astrometry, photometry, and stellar membership probabilities for 56 globular clusters.  They assign a membership probability of 96.1\% to ZNG~1.

\citet{Bond:2021} extracted the coordinates, parallaxes, and proper motions of the entire ZNG sample from the Gaia Early Data Release 3 \citep[EDR3;][]{Gaia_Mission, GaiaEDR3}.  He concluded that the proper motion of ZNG~1 is consistent with cluster membership, but its parallax is discordant.  The Gaia color for this star, $BP - RP = 1.096$, is too red for a star as hot as ZNG~1.  The Gaia Data Release 3 \citep{Gaia:DR3} entry for the star is identical to that of EDR3, but includes estimates of the star's effective temperature and surface gravity.  These values, \teff\ = 5443 K and \logg\ = 2.8, suggest that the star is blended with its G-type companion in the Gaia data.

In Section \ref{sec_rv}, we derive a radial velocity \vhelio\ = $43 \pm 2$ \kms, a value about 2$\sigma$ from the cluster mean.  Based on its proper motion and radial velocity, we conclude that ZNG~1 is a cluster member.

\subsection{Photospheric Abundances}\label{sec_abundance_discussion}

\begin{figure}
\epsscale{1.2}
\plotone{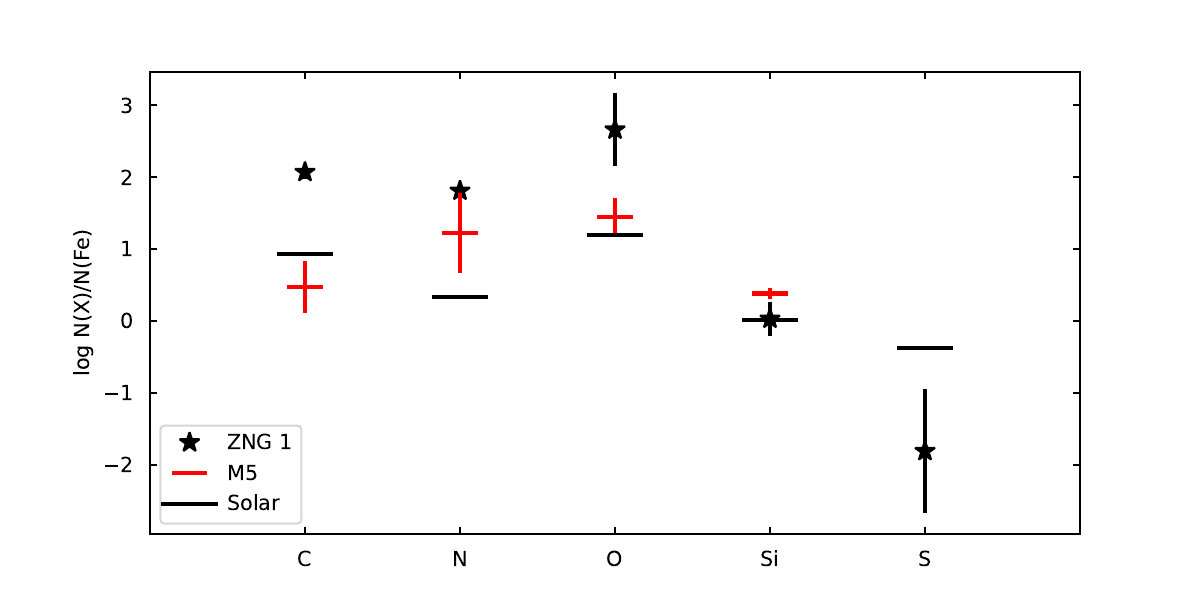}
\caption{Photospheric abundances relative to iron of ZNG~1 (stars), M5 (red lines), and the solar photosphere (black lines).  Cluster values from \citet{Meszaros:2020}.  Solar values from \citet{Asplund:2009}.}
\label{fig_abundance}
\end{figure}

The chemical abundances derived from our line fits are expressed as number fractions relative to helium, which is the dominant species in the star.  Published abundances for the other stars in M5 are expressed in terms of hydrogen.  To facilitate comparison, we assume that iron was neither created nor destroyed in the evolutionary process that produced ZNG~1 and plot $\log N$(X)/$N$(Fe) for each element in \figref{fig_abundance}.  We see that C, N, and O have roughly equal abundances.  C and O are enhanced by about 1.5 dex relative to the cluster, and N is slightly enhanced.  The Si abundance is close to the cluster value, while the S/Fe ratio is subsolar.  Any evolutionary scenario proposed for ZNG~1 must be able to explain this abundance pattern.

\subsection{Stellar Mass and Luminosity}\label{sec_mass}

We can derive a star's radius, and from this its mass and luminosity, by comparing its observed and predicted fluxes.  For ZNG~1, the HUGS catalog quotes Vega magnitudes of 11.97, 12.58, and 14.30 through the F275W, F336W, and F438W filters, respectively.  (We omit the F814W measurement, as it may be contaminated by the star's companion.)  Uncertainties are not provided for these values.  We generate a synthetic spectrum using our best-fit model scaled by a \citet{Fitzpatrick:1999} extinction curve with $R_V = 3.1$ and \ebv\ = 0.03 \citep{Harris:96, Harris:2010}.  For each filter, we compute synthetic stellar magnitudes using the recipe provided in \citet{Riello:2020}.  The ratio between the observed and model fluxes is $\phi$ = (9.20, 9.93, and $10.15) \times 10^{-23}$ for the three filters.  We adopt the mean of these values, $\phi = (9.76 \pm 0.40) \times 10^{-23}$, as our scale factor.

In the synthetic spectra generated by SYNSPEC, the flux is expressed in terms of the flux moment, $H_\lambda$.  If the star's radius and distance are known, then the scale factor required to convert the model spectrum to the flux at earth is $\phi = 4 \pi (R_* / d)^2$ \citep{Kurucz:79}.  Using a combination of Gaia EDR3, \hst, and literature data, \citet{Baumgardt:2021} derive a distance to M5 of $7.479 \pm 0.060$ kpc.  Adopting this distance and our scale factor, we derive a stellar radius  $R_*/R_{\sun}$ of $0.92 \pm 0.02$.  Applying our best-fit surface gravity (\logg\ = 4.47), we find that the stellar mass $M_*/M_{\sun}$ is $0.92 \pm 0.17$.  Finally, combining the stellar radius with our best-fit effective temperature (\teff\ = 43,000 K), we derive a stellar luminosity $\log (L_*/L_{\sun})$ of $3.42 \pm 0.06$.

\subsection{Evolutionary Status}\label{sec_evolution}

AGB stars in globular clusters consist of a C-O core surrounded by concentric shells of He and H.  During most of its time on the AGB, a star is powered by the outer, H-burning shell, the ashes of which rain onto the He shell.  When the mass of the He shell becomes great enough, it ignites in a flash or thermal pulse.  As the He-shell flash dies out, conditions become favorable for the convective dredge-up of material. This so-called third dredge-up brings primary nucleosynthesis products from both H- and He-shell burning to the surface; among these are C, the $s$-process elements, and He \citep{Herwig:2005}.

The effective temperature and luminosity of ZNG~1 place it on the evolutionary tracks of stars evolving from the blue horizontal branch \citep[BHB;][]{Moehler:2019}, yet ZNG~1 is unlikely to be the descendent of an ordinary BHB star.  First, relative to the cluster, the star is equally enhanced in both C and O (\figref{fig_abundance}).  Third dredge-up brings 5 to 10 times more C than O to the stellar surface \citep{Herwig:2005}, so it cannot explain the observed abundance pattern.  Second, the star's mass considerably exceeds that of white dwarfs forming in globular clusters today \citep[$\sim 0.53$ \msun;][]{Kalirai:2009}.  Third, its high rotational velocity cannot be explained by single-star evolution.  Indeed, the high mass and rotational velocity of ZNG~1 strongly suggest that the star is some sort of merger remnant.

While single stars in globular clusters evolve into white dwarfs with C-O cores, binary stars may follow an alternative evolutionary channel: as the more massive star ascends the RGB, it can overflow its Roche lobe, shed its outer envelope, and become a He-core white dwarf (WD) without passing through the HB or AGB phases.  Later, the secondary can undergo the same process, resulting in a pair of He-core WDs.  Given sufficient time, the stars will merge, generally resulting in a helium-rich hot subdwarf of spectral type He-sdO or He-sdB \citep{Jeffery:Zhang:2020}. 

\begin{figure}
\epsscale{1.19}
\plotone{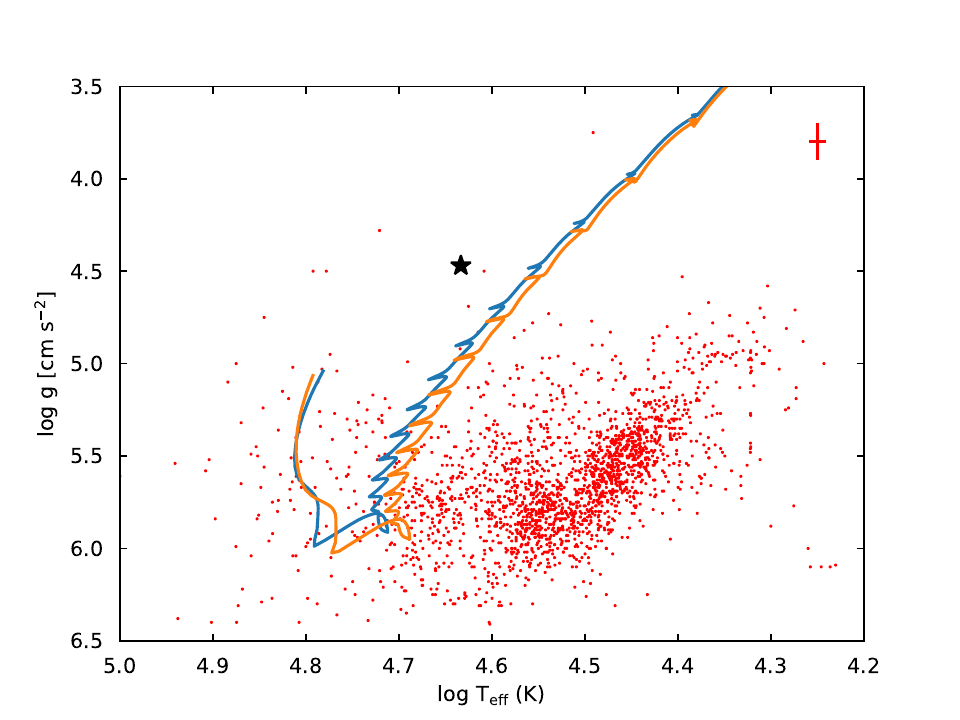}
\caption{Evolutionary tracks from \citet{Zhang:Jeffery:2012} for binary He-WD merger remnants with final masses of 0.8 \msun\ (orange) and 0.9 \msun\ (blue).  Red dots represent the \citet{Geier:2020} sample of helium-rich subdwarf O and B stars.  A representative error bar is plotted in the upper right-hand corner.  The black star marks the effective temperature and surface gravity of ZNG~1.}
\label{fig_jeffery}
\end{figure}

In \figref{fig_jeffery}, we plot evolutionary tracks from \citet{Zhang:Jeffery:2012} for binary He-WD merger remnants with abundances $Z = 0.001$ and masses of 0.8 \msun\ (orange) and 0.9 \msun\ (blue).  These are composite models, in which most of the material of the secondary is transferred directly to the surface of the primary, where it heats up and expands to form a hot corona.  The remainder of the material (0.1 \msun\ in these models) forms a disk around the primary, from which it is transferred to the stellar surface at a rate of $10^{-5}$ \msun\ yr$^{-1}$.  For the 0.9 \msun\ model, the surface gravity rises to \logg\ = 4.5 approximately 65,000 years after the merger.  At an accretion rate of $10^{-5}$ \msun\ yr$^{-1}$, a 0.1 \msun\ disk would last  10,000 years, so would have long since disappeared.  Overplotted are ZNG~1 and a sample of He-sdO and He-sdB stars \citep{Geier:2020}.  ZNG~1 lies in a region of the diagram consistent with a merger product that is evolving toward the helium-burning main sequence.  

\begin{figure}
\epsscale{1.19}
\plotone{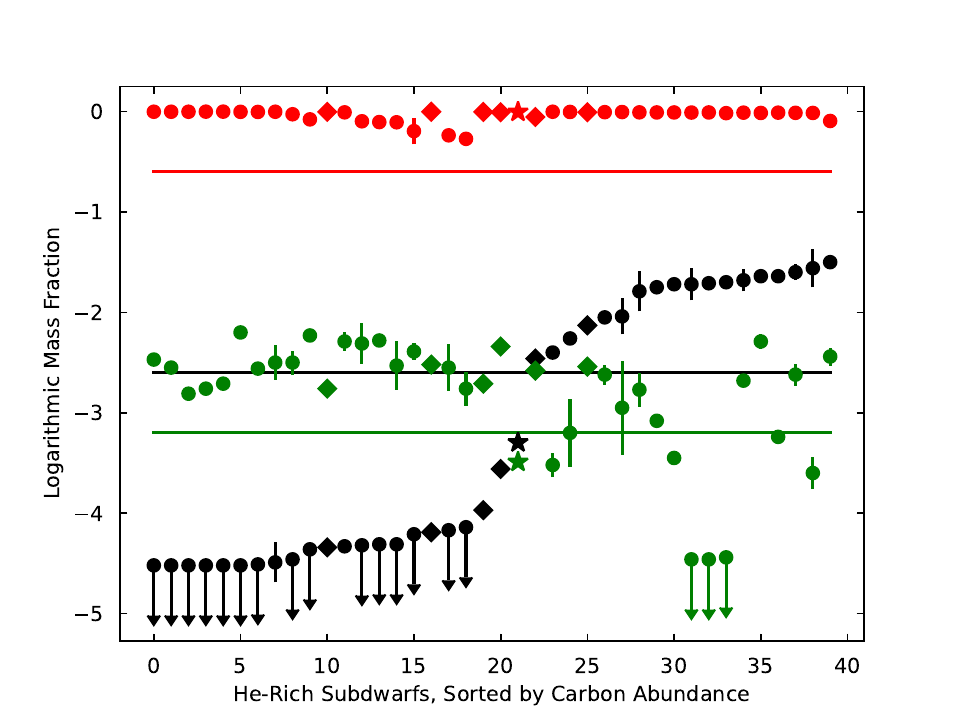}
\caption{The distribution of surface abundance for He-sdO (\citealt{Hirsch:2009}, circles), He-sdB stars (\citealt{Naslim:2010}, diamonds), and ZNG~1 (star), ordered by carbon abundance.  Elements are identified by color: black for C, green for N, and red for He. Horizontal lines indicate the solar abundance of each element.}
\label{fig_hirsch}
\end{figure}

In \figref{fig_hirsch}, derived from similar figures in \citet{Hirsch:2009} and \citet{Zhang:Jeffery:2012}, we plot the distribution of He, C, and N mass fractions in a sample of He-sdO and He-sdB stars.  The C abundance exhibits a bimodal distribution: either C is enhanced above the solar value ($\log \beta_{\mathrm C} > -2.6$) or it is strongly depleted (relative to solar).  The C-rich objects exhibit a wide range of N values ($-3.5 < \log \beta_{\mathrm N} < -2.2$), while the C-poor stars are all N-rich, with N in a narrow range ($-2.8 < \log \beta_{\mathrm N} < -2.0$).   ZNG~1 lies between the C-poor and C-rich groups, though closer to the former, while its N fraction is consistent with those of the C-rich sample.  \citet{Zhang:Jeffery:2012} were able to reproduce the abundance pattern displayed in \figref{fig_hirsch} using composite models with a range of final masses.  The C and N fractions of ZNG~1 correspond to a model with $M \sim 0.67$ \msun, a value slightly lower than our mass estimate.  While the field stars in this figure have solar metallicities, post-merger evolution is driven primarily by a helium-burning shell in the merger remnant, so the tracks and nucleosynthesis should be little affected by the stars' initial metallicity.

Let us consider the high rotational velocity of ZNG~1.   \citet{Gourgouliatos:Jeffery:2006} model the angular-momentum evolution of a merging He-WD binary.  In this model, helium burning in a shell around the core of the primary causes the merger product to expand into a giant.  As the star evolves toward the helium main sequence, it contracts, and its rotational velocity rises dramatically.  If no angular momentum were lost during the merger process, then a 0.7 \msun\ remnant would have a rotational velocity of $\sim 1000$ \kms\ by the time it reached the helium main sequence.  It is likely that  mass---and with it angular momentum---is lost during the merger process, probably at the point when the secondary fills its Roche lobe.  

If high rotational velocities are possible, why are they so unusual?  The rotational velocities of the helium subdwarfs in \figref{fig_hirsch} are less than 35 \kms\ \citep{Hirsch:2009, Naslim:2010}.  Were they more efficient at ejecting angular momentum than ZNG~1?  Perhaps the amount of  angular momentum lost to opacity-driven outflows is a function of metallicity.  Alternatively, the mass of ZNG~1 is relatively high.  Perhaps it inherited a higher angular momentum from its more massive progenitors.

\section{Conclusions}\label{sec_conclusions}
 
We have analyzed archival FUV spectra of the hot UV-bright star ZNG~1 in the globular cluster M5 (NGC~5904).
From these data, we derive the star's heliocentric and rotational velocity, photospheric parameters, and chemical abundances.  The atmosphere is helium-rich and enhanced in CNO (relative to the cluster).  The spectrum exhibits wind features with a terminal velocity near 1500 \kms\ and strong discrete absorptions components (DACs).  The high helium abundance, stellar mass, and rotational velocity suggest that the star is a merger remnant, and its parameters are consistent with models of a pair of merging He-core white dwarfs.

\begin{acknowledgments}

The author wishes to thank M.\ M.\ Miller Bertolami for thoughtful discussions, including the suggestion that ZNG~1 is the remnant of a binary WD merger, and C.\ S.\ Jeffery for providing both his evolutionary tracks and helpful comments on the manuscript.
This work has made use of
NASA's Astrophysics Data System (ADS); % and
the SIMBAD database, operated at CDS, Strasbourg, France; % and
the Mikulski Archive for Space Telescopes (MAST), hosted at the Space Telescope Science Institute, which is operated by the Association of Universities for Research in Astronomy, Inc., under NASA contract NAS5-26555; and
data from the European Space Agency (ESA) mission
{\it Gaia} (\url{https://www.cosmos.esa.int/gaia}), processed by the {\it Gaia}
Data Processing and Analysis Consortium (DPAC,
\url{https://www.cosmos.esa.int/web/gaia/dpac/consortium}). 
Funding for the DPAC has been provided by national institutions, 
in particular the institutions participating in the {\it Gaia} Multilateral Agreement.
Publication of this work is supported by the STScI Director's Discretionary Research Fund.

\end{acknowledgments}


\vspace{5mm}
\facilities{FUSE, HST(STIS)}


%\software{astropy \citep{2013A&A...558A..33A,2018AJ....156..123A},  
%          Cloudy \citep{2013RMxAA..49..137F}, 
%          Source Extractor \citep{1996A&AS..117..393B}
%          }


% \bibliography{apjmnemonic,myref,stars}{}

\begin{thebibliography}{}
\expandafter\ifx\csname natexlab\endcsname\relax\def\natexlab#1{#1}\fi
\providecommand{\url}[1]{\href{#1}{#1}}
\providecommand{\dodoi}[1]{doi:~\href{http://doi.org/#1}{\nolinkurl{#1}}}
\providecommand{\doeprint}[1]{\href{http://ascl.net/#1}{\nolinkurl{http://ascl.net/#1}}}
\providecommand{\doarXiv}[1]{\href{https://arxiv.org/abs/#1}{\nolinkurl{https://arxiv.org/abs/#1}}}

\bibitem[{{Ahmad} {et~al.}(2004){Ahmad}, {Jeffery}, \&
  {Fullerton}}]{Ahmad:2004}
{Ahmad}, A., {Jeffery}, C.~S., \& {Fullerton}, A.~W. 2004, \aap, 418, 275,
  \dodoi{10.1051/0004-6361:20035917}

\bibitem[{{Asplund} {et~al.}(2009){Asplund}, {Grevesse}, {Sauval}, \&
  {Scott}}]{Asplund:2009}
{Asplund}, M., {Grevesse}, N., {Sauval}, A.~J., \& {Scott}, P. 2009, \araa, 47,
  481, \dodoi{10.1146/annurev.astro.46.060407.145222}

\bibitem[{Ayres(2009)}]{Ayres:StarCAT:DOI}
Ayres, T. 2009, HST STIS Echelle Spectral Catalog of Stars (``StarCAT''),
  STScI/MAST, \dodoi{10.17909/T9V01R}

\bibitem[{{Ayres}(2010)}]{Ayres:StarCAT:2010}
{Ayres}, T.~R. 2010, \apjs, 187, 149, \dodoi{10.1088/0067-0049/187/1/149}

\bibitem[{{Baumgardt} \& {Vasiliev}(2021)}]{Baumgardt:2021}
{Baumgardt}, H., \& {Vasiliev}, E. 2021, \mnras, 505, 5957,
  \dodoi{10.1093/mnras/stab1474}

\bibitem[{{Bohlin} {et~al.}(1983){Bohlin}, {Cornett}, {Hill}, {Smith},
  {Stecher}, \& {Sweigart}}]{Bohlin:1983}
{Bohlin}, R.~C., {Cornett}, R.~H., {Hill}, J.~K., {et~al.} 1983, \apjl, 267,
  L89, \dodoi{10.1086/184009}

\bibitem[{{Bond}(2021)}]{Bond:2021}
{Bond}, H.~E. 2021, \aj, 161, 204, \dodoi{10.3847/1538-3881/abe875}

\bibitem[{{de Boer}(1985)}]{deBoer:1985}
{de Boer}, K.~S. 1985, \aap, 142, 321,
\url{https://ui.adsabs.harvard.edu/abs/1985A\&A...142..321D}

\bibitem[{{Dixon} {et~al.}(2004){Dixon}, {Brown}, \& {Landsman}}]{Dixon:2004}
{Dixon}, W.~V., {Brown}, T.~M., \& {Landsman}, W.~B. 2004, \apjl, 600, L43,
  \dodoi{10.1086/381389}

\bibitem[{{Dixon} {et~al.}(2019){Dixon}, {Chayer}, {Reid}, \& {Miller
  Bertolami}}]{Dixon:2019}
{Dixon}, W.~V., {Chayer}, P., {Reid}, I.~N., \& {Miller Bertolami}, M.~M. 2019,
  \aj, 157, 147, \dodoi{10.3847/1538-3881/ab0b40}

\bibitem[{{Dixon} {et~al.}(2007){Dixon}, {Sahnow}, {Barrett},
  {et~al.}}]{Dixon:07}
{Dixon}, W.~V., {Sahnow}, D.~J., {Barrett}, P.~E., {et~al.} 2007, \pasp, 119,
  527, \dodoi{10.1086/518617}

\bibitem[{{Fitzpatrick}(1999)}]{Fitzpatrick:1999}
{Fitzpatrick}, E.~L. 1999, \pasp, 111, 63, \dodoi{10.1086/316293}

\bibitem[{{Gaia Collaboration} {et~al.}(2016){Gaia Collaboration}, {Prusti},
  {de Bruijne}, {Brown}, {Vallenari}, {Babusiaux}, {Bailer-Jones}, {Bastian},
  {Biermann}, {Evans}, \& et~al.}]{Gaia_Mission}
{Gaia Collaboration}, {Prusti}, T., {de Bruijne}, J.~H.~J., {et~al.} 2016,
  \aap, 595, A1, \dodoi{10.1051/0004-6361/201629272}

\bibitem[{{Gaia Collaboration} {et~al.}(2021){Gaia Collaboration}, {Brown},
  {Vallenari}, {Prusti}, {de Bruijne}, {Babusiaux}, {Biermann}, {Creevey},
  {Evans}, {Eyer}, {Hutton}, {Jansen}, {Jordi}, {et~al.}}]{GaiaEDR3}
{Gaia Collaboration}, {Brown}, A.~G.~A., {Vallenari}, A., {et~al.} 2021, \aap,
  649, A1, \dodoi{10.1051/0004-6361/202039657}

\bibitem[{{Gaia Collaboration} {et~al.}(2022){Gaia Collaboration}, {Vallenari},
  {Brown}, {Prusti}, {de Bruijne}, {Arenou}, {Babusiaux}, {Biermann},
  {Creevey}, {Ducourant}, {et~al.}}]{Gaia:DR3}
{Gaia Collaboration}, {Vallenari}, A., {Brown}, A.~G.~A., {et~al.} 2022, arXiv
  e-prints, arXiv:2208.00211, \dodoi{10.48550/arXiv.2208.00211}

\bibitem[{{Geier}(2020)}]{Geier:2020}
{Geier}, S. 2020, \aap, 635, A193, \dodoi{10.1051/0004-6361/202037526}

\bibitem[{{Gourgouliatos} \& {Jeffery}(2006)}]{Gourgouliatos:Jeffery:2006}
{Gourgouliatos}, K.~N., \& {Jeffery}, C.~S. 2006, \mnras, 371, 1381,
  \dodoi{10.1111/j.1365-2966.2006.10780.x}

\bibitem[{{Harris}(1996)}]{Harris:96}
{Harris}, W.~E. 1996, \aj, 112, 1487, \dodoi{10.1086/118116}

\bibitem[{{Harris}(2010)}]{Harris:2010}
---. 2010, arXiv e-prints.
\newblock \doarXiv{1012.3224}

\bibitem[{{Herwig}(2005)}]{Herwig:2005}
{Herwig}, F. 2005, \araa, 43, 435,
  \dodoi{10.1146/annurev.astro.43.072103.150600}

\bibitem[{{Hirsch}(2009)}]{Hirsch:2009}
{Hirsch}, H.~A. 2009, PhD thesis, Friedrich Alexander University of
  Erlangen-Nuremberg, Germany,
\url{https://ui.adsabs.harvard.edu/abs/2009PhDT.......273H}

\bibitem[{{Hubeny}(1988)}]{Hubeny:88}
{Hubeny}, I. 1988, CoPhC, 52, 103, \dodoi{10.1016/0010-4655(88)90177-4}

\bibitem[{{Hubeny} \& {Lanz}(1995)}]{Hubeny:Lanz:95}
{Hubeny}, I., \& {Lanz}, T. 1995, \apj, 439, 875, \dodoi{10.1086/175226}

\bibitem[{{Hubeny} \& {Lanz}(2017)}]{Hubeny:Lanz:2017c}
---. 2017, arXiv e-prints, arXiv:1706.01937.
\newblock \doarXiv{1706.01937}

\bibitem[{{Jeffery} \& {Zhang}(2020)}]{Jeffery:Zhang:2020}
{Jeffery}, C.~S., \& {Zhang}, X. 2020, Journal of Astrophysics and Astronomy,
  41, 48, \dodoi{10.1007/s12036-020-09669-0}

\bibitem[{{Kalirai} {et~al.}(2009){Kalirai}, {Saul Davis}, {Richer},
  {Bergeron}, {Catelan}, {Hansen}, \& {Rich}}]{Kalirai:2009}
{Kalirai}, J.~S., {Saul Davis}, D., {Richer}, H.~B., {et~al.} 2009, \apj, 705,
  408, \dodoi{10.1088/0004-637X/705/1/408}

\bibitem[{{Kimble} {et~al.}(1998){Kimble}, {Woodgate}, {Bowers}, {Kraemer},
  {Kaiser}, {Gull}, {Heap}, {Danks}, {Boggess}, {Green}, {Hutchings},
  {Jenkins}, {Joseph}, {Linsky}, {Maran}, {Moos}, {Roesler}, {Timothy},
  {Weistrop}, {Grady}, {Loiacono}, {Brown}, {Brumfield}, {Content}, {Feinberg},
  {Isaacs}, {Krebs}, {Krueger}, {Melcher}, {Rebar}, {Vitagliano}, {Yagelowich},
  {Meyer}, {Hood}, {Argabright}, {Becker}, {Bottema}, {Breyer}, {Bybee},
  {Christon}, {Delamere}, {Dorn}, {Downey}, {Driggers}, {Ebbets}, {Gallegos},
  {Garner}, {Hetlinger}, {Lettieri}, {Ludtke}, {Michika}, {Nyquist}, {Rose},
  {Stocker}, {Sullivan}, {Van Houten}, {Woodruff}, {Baum}, {Hartig}, {Balzano},
  {Biagetti}, {Blades}, {Bohlin}, {Clampin}, {Doxsey}, {Ferguson},
  {Goudfrooij}, {Hulbert}, {Kutina}, {McGrath}, {Lindler}, {Beck}, {Feggans},
  {Plait}, {Sandoval}, {Hill}, {Collins}, {Cornett}, {Fowler}, {Hill},
  {Landsman}, {Malumuth}, {Standley}, {Blouke}, {Grusczak}, {Reed}, {Robinson},
  {Valenti}, \& {Wolfe}}]{Kimble:STIS:1998}
{Kimble}, R.~A., {Woodgate}, B.~E., {Bowers}, C.~W., {et~al.} 1998, \apjl, 492,
  L83, \dodoi{10.1086/311102}

\bibitem[{{Kurucz}(1979)}]{Kurucz:79}
{Kurucz}, R.~L. 1979, \apjs, 40, 1, \dodoi{10.1086/190589}

\bibitem[{{Lamers} {et~al.}(1987){Lamers}, {Cerruti-Sola}, \&
  {Perinotto}}]{Lamers:87}
{Lamers}, H. J. G. L.~M., {Cerruti-Sola}, M., \& {Perinotto}, M. 1987, \apj,
  314, 726, \url{https://ui.adsabs.harvard.edu/abs/1987ApJ...314..726L}

\bibitem[{{Lanz} {et~al.}(2004){Lanz}, {Brown}, {Sweigart}, {Hubeny}, \&
  {Landsman}}]{Lanz:2004}
{Lanz}, T., {Brown}, T.~M., {Sweigart}, A.~V., {Hubeny}, I., \& {Landsman},
  W.~B. 2004, \apj, 602, 342, \dodoi{10.1086/380904}

\bibitem[{{Lanz} \& {Hubeny}(2003)}]{Lanz:Hubeny:2003}
{Lanz}, T., \& {Hubeny}, I. 2003, \apjs, 146, 417, \dodoi{10.1086/374373}

\bibitem[{{Massa} {et~al.}(2003){Massa}, {Fullerton}, {Sonneborn}, \&
  {Hutchings}}]{Massa:03}
{Massa}, D., {Fullerton}, A.~W., {Sonneborn}, G., \& {Hutchings}, J.~B. 2003,
  \apj, 586, 996, 
\url{https://ui.adsabs.harvard.edu/abs/2003ApJ...586..996M}

\bibitem[{{Massa} {et~al.}(1995){Massa}, {Prinja}, \& {Fullerton}}]{Massa:95}
{Massa}, D., {Prinja}, R.~K., \& {Fullerton}, A.~W. 1995, \apj, 452, 842,
\url{https://ui.adsabs.harvard.edu/abs/1995ApJ...452..842M}

\bibitem[{{M{\'e}sz{\'a}ros} {et~al.}(2020){M{\'e}sz{\'a}ros}, {Masseron},
  {Garc{\'\i}a-Hern{\'a}ndez}, {Allende Prieto}, {Beers}, {Bizyaev},
  {Chojnowski}, {Cohen}, {Cunha}, {Dell'Agli}, {Ebelke},
  {Fern{\'a}ndez-Trincado}, {Frinchaboy}, {Geisler}, {Hasselquist}, {Hearty},
  {Holtzman}, {Johnson}, {Lane}, {Lacerna}, {Longa-Pe{\~n}a}, {Majewski},
  {Martell}, {Minniti}, {Nataf}, {Nidever}, {Pan}, {Schiavon}, {Shetrone},
  {Smith}, {Sobeck}, {Stringfellow}, {Szigeti}, {Tang}, {Wilson}, \&
  {Zamora}}]{Meszaros:2020}
{M{\'e}sz{\'a}ros}, S., {Masseron}, T., {Garc{\'\i}a-Hern{\'a}ndez}, D.~A.,
  {et~al.} 2020, \mnras, 492, 1641, \dodoi{10.1093/mnras/stz3496}

\bibitem[{{Moehler} {et~al.}(2019){Moehler}, {Landsman}, {Lanz}, \& {Miller
  Bertolami}}]{Moehler:2019}
{Moehler}, S., {Landsman}, W.~B., {Lanz}, T., \& {Miller Bertolami}, M.~M.
  2019, \aap, 627, A34, \dodoi{10.1051/0004-6361/201935694}

\bibitem[{{Moos} {et~al.}(2000){Moos}, {Cash}, {Cowie}, {et~al.}}]{Moos:00}
{Moos}, H.~W., {Cash}, W.~C., {Cowie}, L.~L., {et~al.} 2000, \apjl, 538, L1,
  \dodoi{10.1086/312795}

\bibitem[{{Morton}(2003)}]{Morton:03}
{Morton}, D.~C. 2003, \apjs, 149, 205, \dodoi{10.1086/377639}

\bibitem[{{Napiwotzki} \& {Heber}(1997)}]{Napiwotzki:97}
{Napiwotzki}, R., \& {Heber}, U. 1997, in The Third Conference on Faint Blue
  Stars, ed. A.~G.~D. {Philip}, J.~{Liebert}, R.~{Saffer}, \& D.~S. {Hayes},
  441, \url{https://ui.adsabs.harvard.edu/abs/1997fbs..conf..441N}

\bibitem[{{Nardiello} {et~al.}(2018){Nardiello}, {Libralato}, {Piotto},
  {Anderson}, {Bellini}, {Aparicio}, {Bedin}, {Cassisi}, {Granata}, {King},
  {Lucertini}, {Marino}, {Milone}, {Ortolani}, {Platais}, \& {van der
  Marel}}]{Nardiello:2018}
{Nardiello}, D., {Libralato}, M., {Piotto}, G., {et~al.} 2018, \mnras, 481,
  3382, \dodoi{10.1093/mnras/sty2515}

\bibitem[{{Naslim} {et~al.}(2010){Naslim}, {Jeffery}, {Ahmad}, {Behara}, \&
  {{\c{S}}ah{\`\i}n}}]{Naslim:2010}
{Naslim}, N., {Jeffery}, C.~S., {Ahmad}, A., {Behara}, N.~T., \&
  {{\c{S}}ah{\`\i}n}, T. 2010, \mnras, 409, 582,
  \dodoi{10.1111/j.1365-2966.2010.17324.x}

\bibitem[{{Piotto} {et~al.}(2015){Piotto}, {Milone}, {Bedin}, {Anderson},
  {King}, {Marino}, {Nardiello}, {Aparicio}, {Barbuy}, {Bellini}, {Brown},
  {Cassisi}, {Cool}, {Cunial}, {Dalessandro}, {D'Antona}, {Ferraro}, {Hidalgo},
  {Lanzoni}, {Monelli}, {Ortolani}, {Renzini}, {Salaris}, {Sarajedini}, {van
  der Marel}, {Vesperini}, \& {Zoccali}}]{Piotto:2015}
{Piotto}, G., {Milone}, A.~P., {Bedin}, L.~R., {et~al.} 2015, \aj, 149, 91,
  \dodoi{10.1088/0004-6256/149/3/91}

\bibitem[{{Press} {et~al.}(1989){Press}, {Flannery}, {Teukolsky}, \&
  {Vetterling}}]{Press:89}
{Press}, W.~H., {Flannery}, B.~P., {Teukolsky}, S.~A., \& {Vetterling}, W.~T.
  1989, {Numerical Recipes in C: The Art of Scientific Computing} (Cambridge:
  Cambridge Univ. Press), \url{https://ui.adsabs.harvard.edu/abs/1989nrca.book.....P}

\bibitem[{{Prinja} {et~al.}(2002){Prinja}, {Massa}, \& {Fullerton}}]{Prinja:02}
{Prinja}, R.~K., {Massa}, D., \& {Fullerton}, A.~W. 2002, \aap, 388, 587,
\url{https://ui.adsabs.harvard.edu/abs/2002A\&A...388..587P}

\bibitem[{{Rauch}(1993)}]{Rauch:93}
{Rauch}, T. 1993, \aap, 276, 171,
\url{https://ui.adsabs.harvard.edu/abs/1993A\&A...276..171R}

\bibitem[{{Riello} {et~al.}(2021){Riello}, {De Angeli}, {Evans}, {Montegriffo},
  {Carrasco}, {Busso}, {Palaversa}, {Burgess}, {Diener}, {Davidson}, {Rowell},
  {Fabricius}, {Jordi}, {Bellazzini}, {Pancino}, {Harrison}, {Cacciari}, {van
  Leeuwen}, {Hambly}, {Hodgkin}, {Osborne}, {Altavilla}, {Barstow}, {Brown},
  {Castellani}, {Cowell}, {De Luise}, {Gilmore}, {Giuffrida}, {Hidalgo},
  {Holland}, {Marinoni}, {Pagani}, {Piersimoni}, {Pulone}, {Ragaini}, {Rainer},
  {Richards}, {Sanna}, {Walton}, {Weiler}, \& {Yoldas}}]{Riello:2020}
{Riello}, M., {De Angeli}, F., {Evans}, D.~W., {et~al.} 2021, \aap, 649, A3,
  \dodoi{10.1051/0004-6361/202039587}

\bibitem[{{Sahnow} {et~al.}(2000){Sahnow}, {Moos}, {Ake}, {et~al.}}]{Sahnow:00}
{Sahnow}, D.~J., {Moos}, H.~W., {Ake}, T.~B., {et~al.} 2000, \apjl, 538, L7,
  \dodoi{10.1086/312794}

\bibitem[{{Woodgate} {et~al.}(1998){Woodgate}, {Kimble}, {Bowers}, {Kraemer},
  {Kaiser}, {Danks}, {Grady}, {Loiacono}, {Brumfield}, {Feinberg}, {Gull},
  {Heap}, {Maran}, {Lindler}, {Hood}, {Meyer}, {Vanhouten}, {Argabright},
  {Franka}, {Bybee}, {Dorn}, {Bottema}, {Woodruff}, {Michika}, {Sullivan},
  {Hetlinger}, {Ludtke}, {Stocker}, {Delamere}, {Rose}, {Becker}, {Garner},
  {Timothy}, {Blouke}, {Joseph}, {Hartig}, {Green}, {Jenkins}, {Linsky},
  {Hutchings}, {Moos}, {Boggess}, {Roesler}, \&
  {Weistrop}}]{Woodgate:STIS:1998}
{Woodgate}, B.~E., {Kimble}, R.~A., {Bowers}, C.~W., {et~al.} 1998, \pasp, 110,
  1183, \dodoi{10.1086/316243}

\bibitem[{{Zech} {et~al.}(2008){Zech}, {Lehner}, {Howk}, {Dixon}, \&
  {Brown}}]{Zech:2008}
{Zech}, W.~F., {Lehner}, N., {Howk}, J.~C., {Dixon}, W. V.~D., \& {Brown},
  T.~M. 2008, \apj, 679, 460, \dodoi{10.1086/587135}

\bibitem[{{Zhang} \& {Jeffery}(2012)}]{Zhang:Jeffery:2012}
{Zhang}, X., \& {Jeffery}, C.~S. 2012, \mnras, 419, 452,
  \dodoi{10.1111/j.1365-2966.2011.19711.x}

\bibitem[{{Zinn} {et~al.}(1972){Zinn}, {Newell}, \& {Gibson}}]{ZNG:1972}
{Zinn}, R.~J., {Newell}, E.~B., \& {Gibson}, J.~B. 1972, \aap, 18, 390,
\url{https://ui.adsabs.harvard.edu/abs/1972A\&A....18..390Z}

\end{thebibliography}
% \bibliographystyle{aasjournal}



\end{document}